\documentclass[draft]{agujournal}
\draftfalse

\journalname{Space Weather}

\begin{document}

%
%


\title{Ensemble Prediction of a Halo Coronal Mass Ejection Using Heliospheric Imagers}

%
%




\authors{T.~Amerstorfer\affil{1}, C.~M\"ostl\affil{1}, P.~Hess\affil{2}, M.~Temmer\affil{3}, M.L.~Mays\affil{4}, M.~Reiss\affil{3}, P.~Lowrance\affil{5}, Ph.-A.~Bourdin\affil{1}}


\affiliation{1}{Space Research Institute, Austrian Academy of Sciences, 8042 Graz, Austria}
\affiliation{2}{NRC Research Associate, U.S. Naval Research Laboratory Washington, DC 20375, USA}
\affiliation{3}{Institute of Physics, University of Graz, 8010 Graz, Austria}
\affiliation{4}{Heliophysics Science Division, NASA Goddard Space Flight Center, Greenbelt, MD 20771, USA}
\affiliation{5}{IPAC, MS 314-6, California Institute of Technology, 1200 E. California Blvd, Pasadena, CA 91125, USA} 




\correspondingauthor{Tanja Amerstorfer}{tanja.amerstorfer@oeaw.ac.at}




\begin{keypoints}
\item Sun-Earth L1 point is an ideal location for an operational space weather mission carrying HIs
\item Information on the ecliptic extent of a CME improves HI-based predictions
\item ELEvoHI ensemble prediction can be restricted by considering frequency distributions of internal fitting parameters
\end{keypoints}

%
%


\begin{abstract}
The Solar TErrestrial RElations Observatory (STEREO) and its heliospheric imagers (HI) have provided us the possibility to enhance our understanding of the interplanetary propagation of coronal mass ejections (CMEs). HI-based methods are able to forecast arrival times and speeds at any target and use the advantage of tracing a CME's path of propagation up to 1 AU.
In our study we use the ELEvoHI model for CME arrival prediction together with an ensemble approach to derive uncertainties in the modeled arrival time and impact speed. The CME from 3 November 2010 is analyzed by performing 339 model runs that are compared to in situ measurements from lined-up spacecraft MESSENGER and STEREO-B. Remote data from STEREO-B showed the CME as halo event which is comparable to an HI observer situated at L1 and observing an Earth-directed CME. 
A promising and easy approach is found by using the frequency distributions of four ELEvoHI output parameters, drag parameter, background solar wind speed, initial distance and speed.
In this case study, the most frequent values of these outputs lead to the predictions with the smallest errors. Restricting the ensemble to those runs, we are able to reduce the mean absolute arrival time error from $3.5 \pm 2.6$ h to $1.6 \pm 1.1$ h at 1 AU. 
Our study suggests that L1 may provide a sufficient vantage point for an Earth-directed CME, when observed by HI, and that ensemble modeling could be a feasible approach to use ELEvoHI operationally.
\end{abstract}

\section{Introduction}


Coronal mass ejections (CMEs) are the drivers of the most intense geomagnetic storms at Earth. The composition of enhanced particle density, high speed and an enclosed magnetic flux rope with an increased magnetic field strength can lead to severe disturbances on Earth and the difficulties with predicting these phenomena are currently fueling world-wide efforts to better understand and forecast them.
In the last decade NASA's twin satellites the Solar TERrestrial RElations Observatory (STEREO) have facilitated a deep insight into the interplanetary propagation of coronal mass ejections (CMEs). In particular, the wide-angle heliospheric imagers (HI) enabled the development of a multitude of methods for analyzing the evolution of CMEs through interplanetary (IP) space \citep[][]{kahweb07, she99, rou08,lug09b,moe11,dav12,moedav13,rol12,rol13,rol14}. A recent review on HI and according methods can be found in \citet{har17}. Case studies using HI-based prediction models assuming constant propagation speed find an arrival time error of about $8 \pm 6$ h, arrival speeds are mostly overestimated by some $100$ km s$^{-1}$ \citep[e.g.][]{moe14}. Using the WSA-ENLIL+Cone model, i.e.\ the Wang-Sheeley-Arge coronal model \citep[WSA;][]{argpiz00,arg04} combined with the ENLIL solar wind model \citep[][]{ods03}, for predicting the arrival of 17 events, \citet{may15} applied an ensemble approach and found a mean absolute arrival time error of $12.3$ h, which is in the same range as other studies show \citep[][]{mil13,vrs14,rol16}. Studies covering larger samples of events are rare but reflect a realistic picture of CME arrival time prediction. In a recent study by \textit{Wold et al.} (accepted for publication by the Journal of Space Weather and Space Climate), almost seven years of operational CME arrival predictions using the WSA-ENLIL+Cone model were assessed. During this period, 273 events were predicted and observed at Earth with a mean absolute arrival time error of $10 \pm 0.9$ h. That study represents the currently achieved arrival time error when predicting CMEs at Earth as the WSA-ENLIL+Cone model is the state-of-the-art and widely used for operational space weather forecasting.
In \citet{tuc15} 60 CME arrival predictions were performed, resulting in an absolute average error in transit time of 19 h. In that study, STEREO HI beacon data were used, which depicts the situation of operational forecasts when using HI near real time data. Another recent study by \citet{moe17} predicted the arrival of 1337 CMEs based on HI science data from eight years of observations and used the self-similar expansion fitting method \citep[][]{moedav13}. From this dataset, 315 CMEs were deteced in situ. Assuming a constant propagation speed, a mean absolute arrival time error of $14.2$ h was found. It is expected that the arrival time error can be reduced when the interaction with the ambient medium is taken into account.
Currently, for operational forecasting mainly coronagraph observations from LASCO onboard the Solar and Heliospheric Observatory (SoHO) are used. These observations have two main handicaps compared to HI observations. First, SoHO is located at the Lagrangian L1 point, situated about $1.5$ million km in front of Earth along the Sun-Earth line. This provides a head-on vantage point of Earth-directed CMEs, which appear as halo CMEs in such observations. The expansion of such halo CMEs is an indicator for the propagation speed \citep[][]{schw05} and can be used to forecast the arrival time at Earth. Second, LASCO C3 observes the space around the Sun up to $30$ solar radii (R$_\odot$), which corresponds to only $15$\% of the Sun-Earth distance. From STEREO HI observations we know that the interplanetary propagation of CMEs is far from being undisturbed. Therefore, it is of high value to be able to follow a CME's evolution along a larger distance than coronagraphs provide \citep[e.g.][]{col13}. 
Besides improving the prediction of arrival time and speed of a CME at Earth, there is an even more important issue, namely to reduce the false alarm rate, which is the percentage of CMEs predicted to impact Earth that actually miss. \citet{may15} indicate the false alarm rate to be $38$\% when predicting CME arrivals using the WSA-ENLIL+Cone model.
CMEs can be strongly influenced by different phenomena in the solar wind like other CMEs or the background solar wind itself. Besides the typical deceleration or acceleration of fast or slow events, they can be forced to change their overall direction of motion due to the influence of magnetic forces close to the Sun \citep[][]{kayoph15,moe15} or due to other CMEs farther out in IP space \citep[e.g.][]{lug12}.

The CME studied in this article erupted on 3 November 2010, associated with a C4.9 flare close to the eastern limb of the Sun peaking at 12:21 UT \citep[][]{ree11}. Various studies analyzed the eruption consistent with the classical standard flare-CME model. 
\citet[][]{bai12} studied the metric type \textrm{II} burst, which was associated with the eruption. The authors found that the burst was located ahead of the hot core of the erupting plasmoid, which is an indication for a piston-driven shock. \citet[][]{zim12} analyzed the same event in more detail and came to a similar conclusion, namely the presence of a piston-driven shock. They note that the shock wave may have transformed to a freely propagating blast wave during its evolution. However, \citet{kuminn13} discovered fast waves at the onset of the flare, which hints at the type \textrm{II} burst being caused by a blast wave rather than by a piston-driven shock. Due to the exceptional good observations in extreme ultraviolet, a multitude of studies investigated the multi-thermal dynamics and the early stage of the eruption \citep[e.g.][]{chen11,fou11,hankon13}.

In this study, we aim to test the L1 point as a possible location for an operational heliospheric imager to monitor Earth-directed CMEs. We use the advantage of the CME on 3 November 2010, directed towards STEREO-B and observed remotely by HI as well as in situ by the same spacecraft, to simulate the situation of an Earth-directed CME observed from L1. Additionally, the CME was detected in situ by the MESSENGER spacecraft, which was almost exactly lined-up with STEREO-B during the time of the event. Ensemble predictions (339 model runs) from the current state-of-the-art HI elongation fitting method ELEvoHI \citep[][]{rol16} as well as constraining the predictions with additional information on the CME mass and on the frequency distribution of four ELEvoHI output parameters show a promising new possibility for more accurate CME arrival predictions.

\section{Event Overview and Data}

\subsection{Remote Observations}

\begin{figure}[h]
\centering
\includegraphics[width=\textwidth]{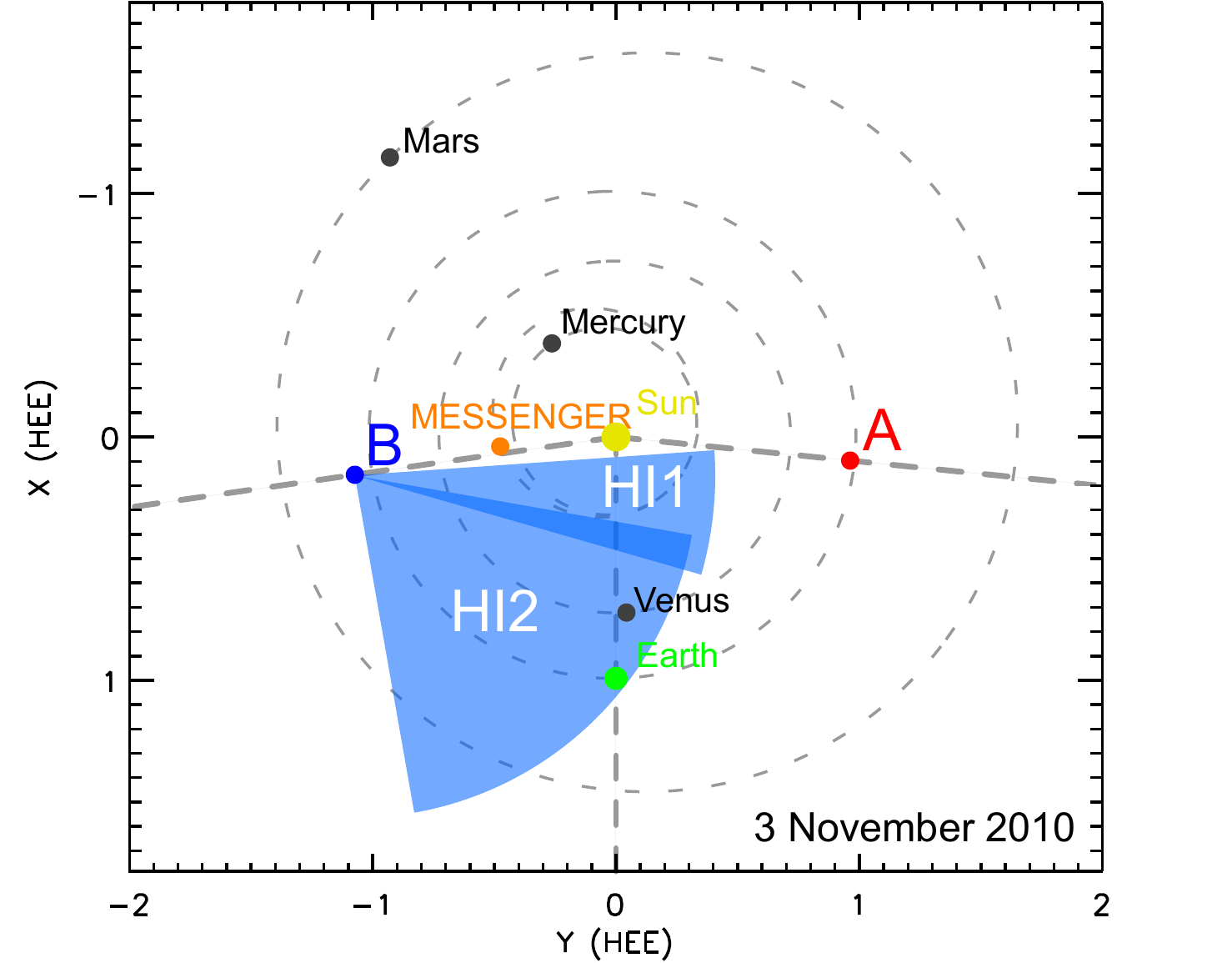}
\caption{Positions of STEREO, MESSENGER and the planets of the inner heliosphere at the time of launch of the CME under study. MESSENGER and STEREO-B were radially aligned, both spacecraft measured the CME in situ. The fields of view of HI1-B and HI2-B are marked by the blue areas. For this study only HI data from HI1 are used.}
\label{fig:innerhel}
\end{figure}

CMEs are commonly observed by coronagraphs, where the bright photospheric light is shielded by occulter disks. This enables the observation of the faint solar corona. Situated at the L1 point, the Solar and Heliospheric Observatory (SoHO) carries two of such instruments, LASCO C2 and C3 \citep[][]{bru95}, having a field of view of 2 to 6 R$_\odot$ and 3.7 to 30 R$_\odot$, respectively. 
The STEREO mission was launched in 2006 and consists of two twin satellites, STEREO-A(head) and STEREO-B(ehind), both equipped with the same set of instruments. Part of STEREO's SECCHI suite \citep[][]{how08} are two coronagraphs, COR1 and COR2, observing an area of 1.4 to 4 R$_\odot$ and 2 to 15 R$_\odot$ around the Sun. At the time of the CME event under study, STEREO-A was 84$^\circ$ ahead of Earth, STEREO-B was 82$^\circ$ behind, i.e.\ they were separated by 166$^\circ$.
In addition to coronagraph observations we use data from the heliospheric imagers, HI1 and HI2, wide-angle white light cameras observing the space between the Sun and 1~AU. HI1 has a field of view of 4 to 24$^\circ$ elongation (the angle between the Sun-spacecraft line and the line of sight), HI2 observes an area of 18 to 88$^\circ$ elongation. Figure \ref{fig:innerhel} shows the positions of STEREO, MESSENGER and the planets of the inner heliosphere. The blue shaded areas mark the fields of view of HI1-B and HI2-B. In this study only HI data from HI1-B are used for the ELEvoHI arrival predictions.

The CME under study was first observed by LASCO C2 on 3 November 2010 at 12:36~UT and  entered the field of view of C3 at 14:06~UT. In STEREO-B COR2 the CME was visible as a halo, while in STEREO-A COR2 it appeared as backside halo CME. It entered the field of view of STEREO-B HI1 on 4 November 2010 at 4:49~UT and the STEREO-A HI1 field of view at 3:29~UT. In STEREO-A and B HI2 the CME was first visible on 5 November 2010 at 10:10~UT.

\subsection{In Situ Observations}

\begin{figure}[h]
\centering
\includegraphics[width=0.7\textwidth]{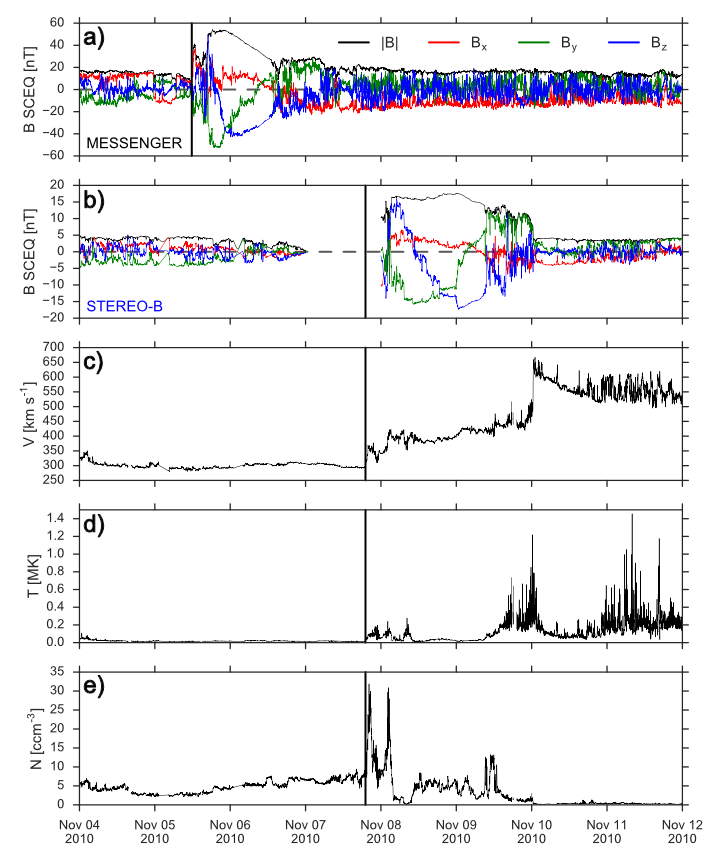}
\caption{In situ observations by MESSENGER and STEREO-B of the early November 2010 ICME. Panel a) shows the magnetic field components ($B_x$ red, $B_y$ green, $B_z$ blue) and total field (black) at MESSENGER in the spacecraft equatorial coordinate system, which is similar to Heliocentric Earth Equatorial (HEEQ) coordinates except that the system is centered on the spacecraft, not Earth. The vertical solid lines indicate the arrival of  a shock.
Panel b) shows the magnetic field at STEREO-B in a similar format. The proton bulk speed at STEREO-B is shown in panel c), panel d) displays the proton temperature and e) the density. The shock arrival time at STEREO-B at the solid vertical line is derived from the plasma parameters as the magnetic field has a data gap.}
\label{fig:insitu}
\end{figure}

The first detection of the CME shock was recorded on 5 November 2010 at 11:46~UT by the MESSENGER spacecraft, which was situated at $0.48$~AU $84^\circ$ east of Earth.
During its cruise phase between August 2004 and March 2011, the magnetometer onboard MESSENGER \citep[MAG;][]{and07} measured the interplanetary magnetic field vector in the solar wind. Figure \ref{fig:insitu}a shows the magnetic field vector in SpaceCraft Equatorial Coordinates (SCEQ) with red, green and blue lines being the $x$, $y$, $z$ components and the black line being the total magnetic field. In the SCEQ coordinate system, the $z$-axis is the solar rotation axis, the $x$-axis points from the Sun to the spacecraft and $y$ completes the right-handed triad, pointing to solar west. The NES-type flux rope started at 16:53~UT and ended at 13:24~UT, having a right-handed chirality and an axis orientation with a low inclination relative to the ecliptic plane \citep[][]{botsch98}.

At 7 November at 19:05~UT the CME shock arrival was detected by STEREO-B, located at 1.08~AU and $82^\circ$ east of Earth. In contrast to MESSENGER, STEREO also provides plasma measurements. The CME sheath region arrives with a speed of $\approx 350$ km s$^{-1}$, while during the flux rope interval, the speed is $\approx 400$ km s$^{-1}$ during its first half and increases to more than $450$ km s$^{-1}$ (Figure \ref{fig:insitu}b,c). The reason for this speed increase seems to be the high-speed solar wind stream, which is pushing the magnetic flux rope from behind. The result of this interaction is a reverse shock behind the flux rope with a speed of $\approx 600$ km s$^{-1}$. The magnetic flux rope started on 8 November at 03:28~UT and lasted until 9 November at 09:11~UT. Similar to the magnetic signature at MESSENGER, we find a low inclined flux rope with a positive chirality, so the overall flux rope structure has not changed. 
Usually, CMEs expand during their interplanetary propagation, which increases their duration and decreases their magnetic field strength. This event is no exception, as the mean magnetic field strength in the magnetic flux rope has decreased by a factor of 2.7 from 43.6 $\pm$ 8.6 nT at MESSENGER to 16.2 $\pm$ 0.9 nT at STEREO-B. The duration of the flux rope at STEREO-B (30.3 h) is 50~\% larger than the duration at MESSENGER (20.2 h).

During the CME impact the Spitzer Space Telescope was located about $34^\circ$ west of STEREO-B, which provides an opportunity to further test ELEvoHI with an additional in situ detection from a third spacecraft. A good indication for space weather events at Spitzer is the number soft scrub errors, which can be directly related to solar flares or CMEs \citep[][]{chen14}, but no increase in these errors was observed. Since Spitzer is mostly affected by high-energy particle hits, i.e.\ energies of about 100 MeV and from the High Energy Telescopes from STEREO-B/IMPACT \citep[][]{luh08,vonros08} we conclude that no high energetic particles have been observed during the time of arrival of the CME. 

\section{Methods}

\subsection{ELEvoHI}
\label{sec:ELEvoHI}

ELEvoHI is a CME prediction utility first presented in \cite{rol16}, where it was applied to 21 CMEs observed by HI. It uses HI observations as input and predicts arrival times and speeds at the target of interest. ELEvoHI combines various methods, which have been already used to investigate the interplanetary evolution of CMEs. One part of this prediction tool is the Elliptic Conversion method \citep[ElCon;][]{rol16}, which converts the observed HI elongation angle into a unit of distance, i.e.\ it reveals the interplanetary CME kinematics \citep[e.g.][]{bar17} including the initial speed, $v_{\mathrm init}$, for the prediction. For the event under study, the mean $v_{\mathrm init}$ for the whole ensemble is $v_{\mathrm init}=541 \pm 42$ km s$^{-1}$. ElCon provides the possibility to modify the extension of the CME shape within the ecliptic plane as suitable for each event under study. Depending on the geometry of the run (as we vary the shape for each of the 339 predictions), $v_{\mathrm init}$ varies between 460 and 660 km s$^{-1}$. Besides the direction of motion, $\phi$, the aspect ratio of the ellipse semiaxes, $f$, and the angular half width, $\lambda$, can be fixed, each of the three parameters is assumed to stay constant during propagation. For the equations used by ElCon, we refer to the Appendix section in \citet{rol16}. The next technique implemented within ELEvoHI is the numerical fitting (downhill simplex method) of the ElCon time-distance profile using a drag-based equation of motion \citep[][]{vrs13}. Here, it is assumed that the propagation of a CME is exclusively dominated by the drag force exerted by the solar wind: 

\begin{equation}
R_{\rm DBM}(t) = \pm \frac{1}{\gamma} \ln [1 \pm \gamma (v_{\rm init}-w) t] + w t + r_{\rm init},
\label{eq:dbm}
\end{equation}
where $R_{\rm DBM}(t)$ is the radial distance from Sun-center in R$_\odot$, $\gamma$ is the drag parameter, which is usually ranging between $0.2\times10^{-7}$ km$^{-1}$ and $2\times10^{-7}$ km$^{-1}$. $v_{\rm init}$ and $r_{\rm init}$ are the initial speed and distance, respectively, and $w$ is the background solar wind speed. The drag parameter is defined as

\begin{equation}
\gamma=c_{\rm D} \frac{A \rho_{\rm sw}}{m_{\rm CME}},
\label{eq:gamma}
\end{equation}
with $c_{\rm D}$ being a dimensionless drag coefficient (assumed to be 1), $A$ is the CME cross-section the drag is acting on, $\rho_{\rm sw}$ is the solar wind density and $m_{\rm CME}$ is the CME mass.
$r_{\rm init}$ as well as the end point of the fit usually need to be defined manually. In this study, the ElCon time-distance profile is fitted between $\approx 30-100$ R$_\odot$, i.e.\ only HI1 data were needed for prediction. The sign $\pm$ is positive when $v_{\rm init} > w$ and negative when $v_{\rm init} < w$. To find the most adequate value for $w$, ELEvoHI reads in in situ data from 1~AU from the same time range as the HI observations and performs several fits with different values for $w$. The fit with the smallest residuals reveals the background solar wind speed. We note that this approach is suitable for real time prediction since both kinds of data (HI as well as in situ solar wind speed from 1~AU) are available in (near) real time. Another approach of DBM fitting is presented by \citet{zic15}, who iteratively fit a time-distance profile using successive input from HI.
The last step of ELEvoHI is to perform the prediction. This is done by the Ellipse Evolution model \citep[ElEvo;][]{moe15}, which uses the information gained by ElCon and drag-based fitting as input. ElEvo as well assumes an elliptical shape for the CME front and runs the drag-based model \citep[][]{vrs13} to perform the prediction.

\subsection{CME mass determination}

\begin{figure}[h]
\centering
\includegraphics[width=0.8\textwidth]{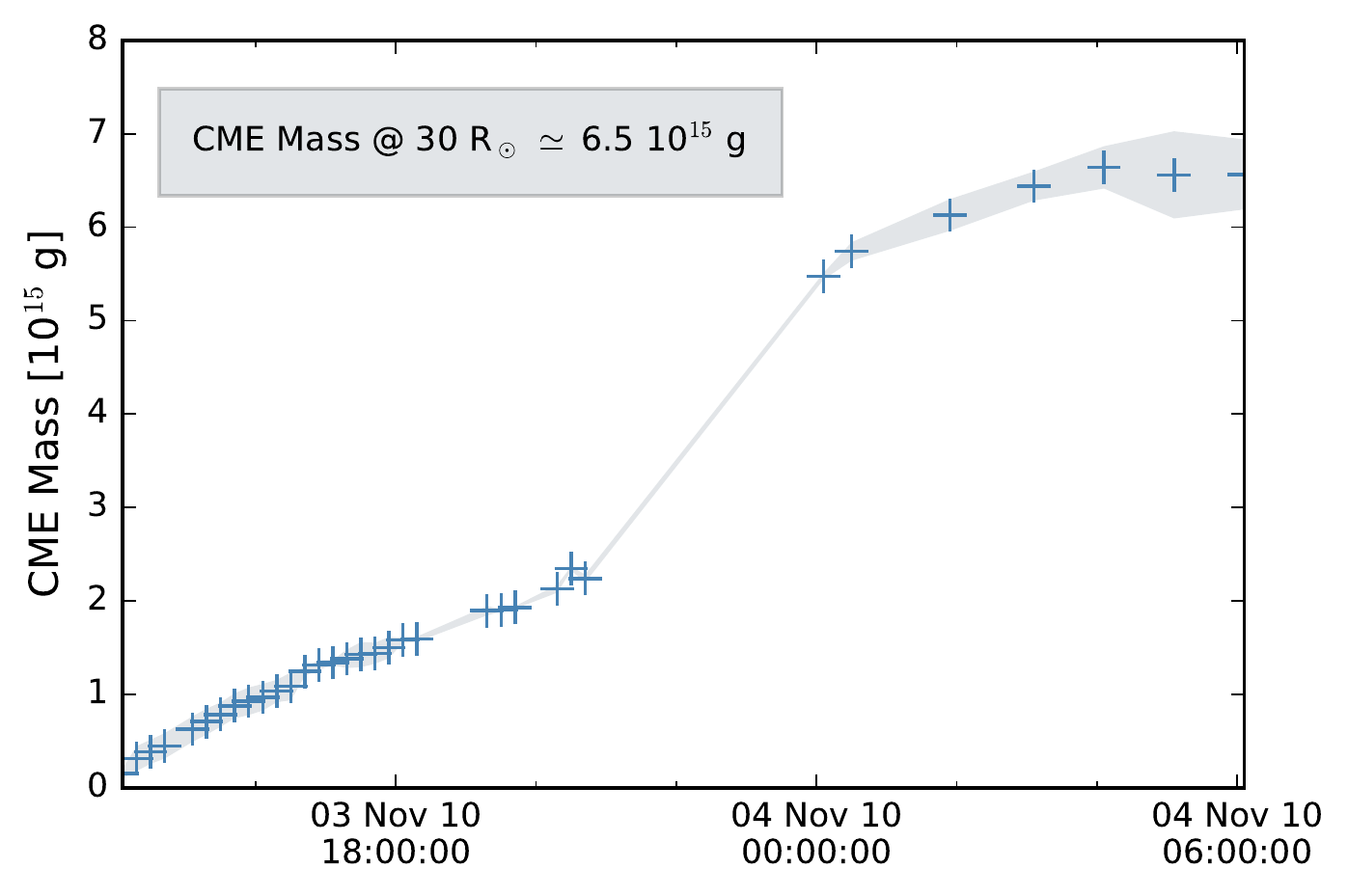}
\caption{CME mass evolution versus time derived from LASCO C3 white light imagery covering the distance range $\sim$5--30 R$_\odot$. The final mass is derived as average over the last three data points, between 4--6 UT on 4 November 2010.}
\label{fig:mass}
\end{figure}

CMEs can be observed in white-light as photons are scattered off the coronal electrons which build the CME structure. Assuming that the CME lies in the plane of sky, we derive the CME mass evolution using the excess brightness as measured from white-light data. LASCO C3 data preparation was done to correct for instrumental effects and calibrate in units of mean solar brightness. To derive the excess brightness a pre-event image is subtracted \citep[see e.g.][]{vou00}. Assuming that the ejected CME material consists of completely ionized hydrogen (90\%H) and helium (10\%He) the mass is calculated using the Thomson scattering function by \citet{bil66}. As shown in Figure~\ref{fig:mass}, the CME mass evolves very slowly over several hours, before a strong increase is observed. This can be interpreted as a slow streamer-blowout CME. Since we describe in the beginning the type II burst related to the CME and estimated speeds of the order of 1500 km s$^{-1}$ this might need some additional explanation:
Though the CME started very impulsively and produced a type II burst \citep[e.g.][]{bai12}, the further evolution is rather moderate and the POS speed over LASCO field of view yields about 250 km s$^{-1}$. The rapid deceleration, deviation from radial propagation, and slow increase in mass would suggest that the CME might have interacted with a streamer, resulting in its blowout \citep[e.g.][]{ese15}. However, here we have to note that the initial speed at $\approx 30$ R$_\odot$ derived from HI observations lies in a range of 490--570 km s$^{-1}$, which seems to be more reliable than the speed derived from coronagraph observations as the CME arrived with 350--400 km s$^{-1}$ at 1~AU.
The final mass for the fully developed CME, as observed in LASCO/C3 close to the outermost boundary of C3 FoV at 30 R$_\odot$, is derived over the time range from 4 to 6~UT on 4 November 2010 (last three data points in Figure~\ref{fig:mass}) with m$_{30}\approx 6.5~10^{15}$~g.

\subsection{Graduated Cylindrical Shell fitting}

To determine the CME geometry in the corona, the Graduated Cylindrical Shell (GCS) forward modeling technique \citep{the06, the09} is employed. This model reduces the CME magnetic ejecta, i.e.\ the flux rope, to a function of six free parameters: the propagation longitude and latitude, the tilt angle of the CME central axis, the separation width of the CME legs, the aspect ratio between the CME major and minor cross sections, and the height of the CME nose at a particular time. These parameters are determined by utilizing approximately co-temporal images from SECCHI and LASCO to fit the proscribed geometry to what is observed from multiple viewpoints at different times. As in \citet{hes15}, most parameters are kept as fixed as possible to provide a unique solution to the CME geometry. However, while that study focused only on fast CMEs, the slower speed of the 3 November 2010 CME required a slight adjustment to the longitude with time to account for solar rotation. 


The GCS fitting parameters used for this CME included a latitude of $-2.24^\circ$, a tilt angle of $16.77^\circ$, a half angle of $38.57^\circ$ and an aspect ratio of $0.29$. These parameters were fixed throughout the propagation. The Carrington longitude was gradually changed from $219^\circ$ to $209^\circ$. The height of the nose was $13.43$ R$_{\odot}$ at 00:54 UT on 4 November and the last measurement performed in HI1 had a height of $93.28$ R$_{\odot}$ at 06:09 UT on the 5 November.
Observationally, there appears to be a coherent flux rope structure in the coronagraph data that serves as the basis for these fits. When processing the data with a running difference, another structure is visibile, which can be a sign of a CME driven shock \citep{hes14}. Regardless of the source of this structure, it did provide a complication in determining the exact extent of the CME width. 
In order to try and get a sense of the possible error of the event, an extremely wide fit was performed to include this structure. Most of the parameters are similar to the original fit performed, but the aspect ratio was increased to $0.36$ and the half-angle width was $57.58^\circ$. This CME is almost certainly too wide, but it may be a better fit to the entire density structure that is visible, especially in HI-1.

The GCS model provides a 3-dimensional geometry in the corona. To generate the inputs for EIEvoHI, the extent of the CME leading edge in the ecliptic plane must be determined. As first presented by \citet{col13}, this can be done analytically utilizing the detailed geometry of the model presented in \citet{the11}. If the CME is propagating well away from the ecliptic or has a significant tilt, the ecliptic cut of the GCS geometry will vary more significantly due to slight changes. However, for a CME with a central axis that is close the ecliptic plane, this will be less significant.

\section{Ensemble of ELEvoHI predictions}

\subsection{Determine the CME shape and direction}

\begin{figure}[h]
\centering
\includegraphics[width=\textwidth]{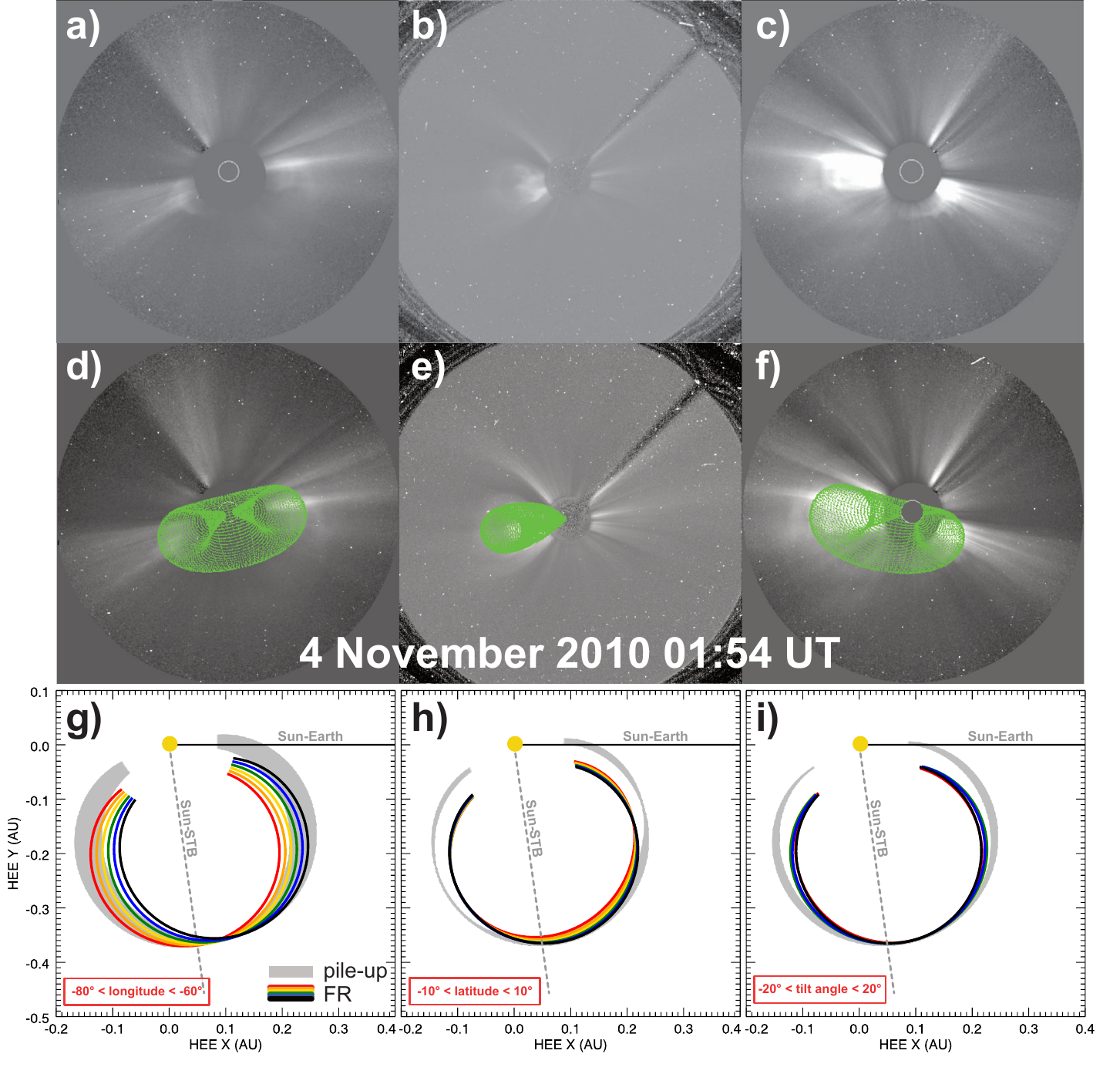}
\caption{GCS modeling applied to COR2-B (a, d), LASCO/C3 (b, e), and COR2-A (c, f) base difference images and shape variations of the eclitpic cut of the GCS shape when varying the tilt angle (g), the longitude (h) and the latitude (i) within their error ranges. The colored ellipses stem from GCS fitting to the ejecta, i.e.\ the flux rope, the gray area shows the variation of the ecliptic cut from GCS fitting to the CME density pile-up.}
\label{fig:gcs_cut}
\end{figure}

From the cut of the GCS fit with the ecliptic plane we measure the input parameters for the CME shape needed by ELEvoHI, i.e.\ the propagation direction, $\phi$, the inverse ellipse aspect ratio of the semiaxes, $f=b/a$, and the angular half width, $\lambda$. 
Panels d), e) and f) of Figure \ref{fig:gcs_cut} show the GCS shape overlaid on base difference images (a)--c)) produced from observations of COR2-B, LASCO C3 and COR2-A. The lowest panels of Figure \ref{fig:gcs_cut} show the variation of the ecliptic cut when the GCS longitude (g), the latitude (h) or the tilt angle (i) are varied within estimated errors of the GCS model ($-80^\circ \leq$ longitude $\leq -60^\circ$, $-10^\circ \leq$ latitude $\leq -10^\circ$, $-20^\circ \leq$ tilt angle $\leq 20^\circ$) leading to the possible range of the ELEvoHI input parameters. A full examination of the errors in the GCS model has not been undertaken, but based on experience with the model and the cross-comparison of fits between various individuals, we believe these values are reasonable and conservative. A change in the longitude is the most obvious in the ecliptic cut for the 3 November 2010 CME as it controls the pointing of the nose of the CME. Because this CME is low tilt and from a near equatorial latitude (in coronagraph observations), varying those parameters has very little effect on the shape of the CME. Even the effect of the longitude is not likely to affect the results near the CME nose, but may impact the ability to determine the exact extent of the longitudes that will or will not be impacted by the CME flank, and therefore may be a source of both missed detections and false alarms for CMEs that propagate farther from the Sun-Earth line. The gray areas in the lower panels of Figure \ref{fig:gcs_cut} mark the attempt to fit the GCS model rather to the dense area surrounding the ejecta. This approach might be more consistent with other assumptions of ELEvoHI, especially when tracking a CME in HI at its shock front and not at its cavity. Taking into account the variations of the GCS fit to the dense CME parts as well as to the ejecta we find the following range of the ELEvoHI input parameters: $2^\circ \leq \phi \leq 14^\circ$, $0.76 \leq f \leq 1$, and $55^\circ \leq \lambda \leq 85^\circ$, having steps of $\Delta \phi=2^\circ$, $\Delta f = 0.04$, and $\Delta \lambda = 5^\circ$. Within these boundaries we perform $N=343$ runs for the input triplets $\{\phi, f, \lambda \}$ with $n_\phi=7$, $n_f=7$, and $n_\lambda=7$, i.e. every possible combination of the three input parameters is part of the ensemble. For the triplets $\{12^\circ, 0.76, 85^\circ\}$, $\{14^\circ, 0.76, 85^\circ\}$, $\{14^\circ, 0.76, 75^\circ\}$, $\{14^\circ, 0.76, 80^\circ\}$ no solutions exist, i.e.\ the DBM fits do not converge and the total number of runs reduces to $N=339$.

\subsection{ELEvoHI forecast}

\begin{figure}[h]
\centering
\includegraphics[width=0.85\textwidth]{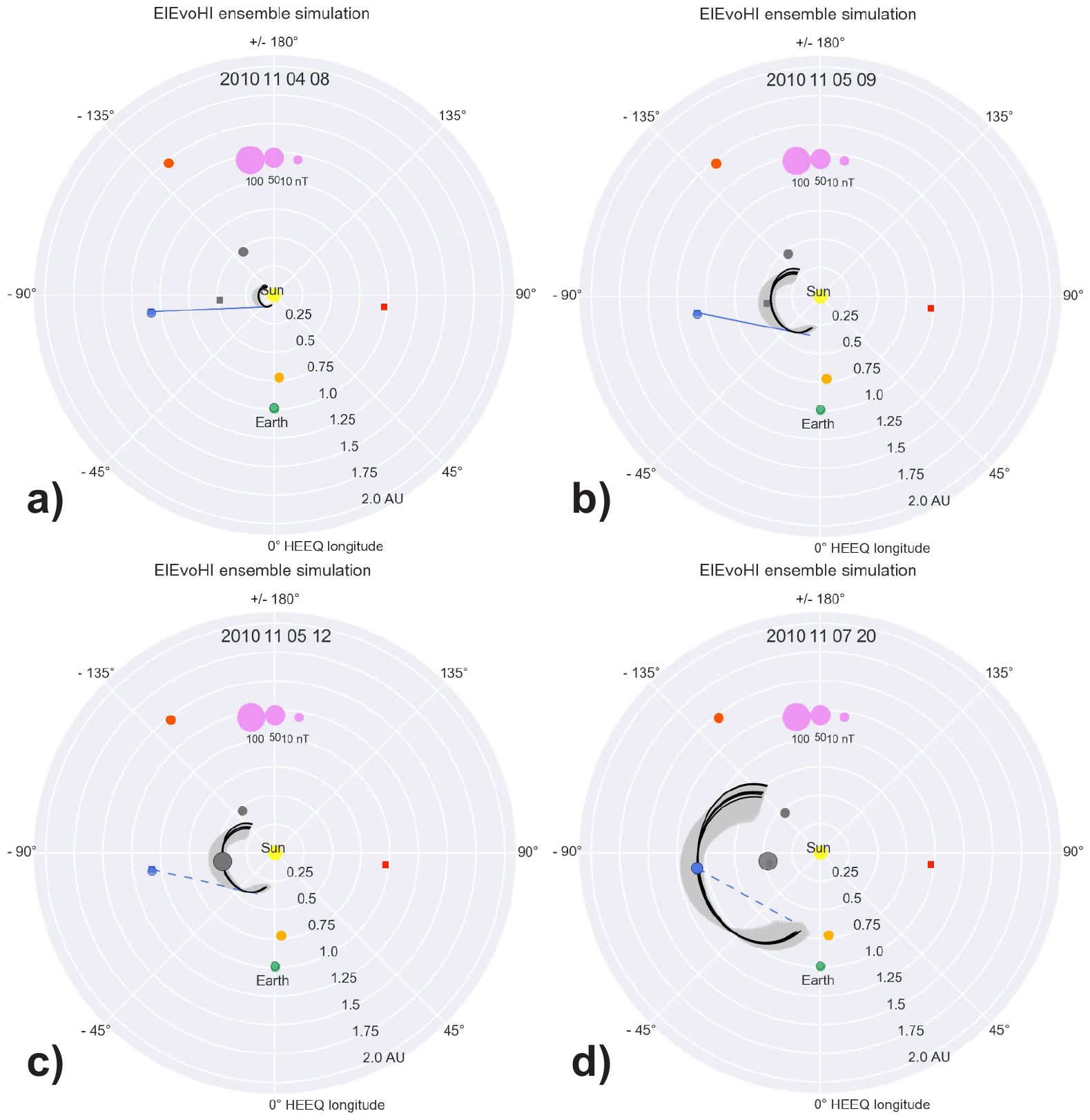}
\caption{Visualization of ELEvoHI results. The black curves correspond to the CME shapes leading to the best prediction at MESSENGER and STEREO-B, respectively. The dark gray area is the entity of all other runs from the ensemble. In panel a) and b) the blue tangent corresponds to the first and last elongation measurement used for the prediction. These elongations correspond to about 30 and 100 R$_\odot$. In panel c) the CME arrives at MESSENGER, the dashed tangent proofs the consistency with HI observations, which are not taken into account for the predictions anymore. The size of the filled circle at the location of MESSENGER marks the magnetic field strength measured in situ. In panel d) the CME impacts STEREO-B, the dashed line marks the last HI observation. An animated version of this figure is available online.}
\label{fig:elevohi}
\end{figure}

Figure \ref{fig:elevohi} shows four different time steps of the ELEvoHI ensemble prediction. In each panel, the Sun is in the center, MESSENGER is marked as a gray square, STEREO-B is marked as a blue filled circle. Panel a) shows the time of the first HI elongation measurement (blue tangent) used as input, panel b) shows the time of the last HI elongation measurement used, the black ellipses correspond to those runs with an arrival time within $\pm 0.5$ h at MESSENGER and STEREO-B, respectively. The dark gray area is the whole ensemble. Panel c) shows the time of the in situ arrival at MESSENGER, the dashed tangent shows the HI elongation measurement from the same time, still consistent with the model output, but not used for calculation anymore. Panel d) shows the CME impact at STEREO-B, the dashed blue line marks the last elongation measurement. An animated version of Figure \ref{fig:elevohi} is available online.

From the 339 predictions performed (runtime for the whole ensemble prediction is less than one hour on a desktop computer), 50 lead to an arrival time error of less than $\pm 1$ h at MESSENGER as well as at STEREO-B. Reducing the arrival time window to $\pm 0.5$ h results in 22 events. 83 \% of the predictions lie within $\pm 6$ h. The best arrival time prediction yield the triplets $\{2^\circ, 0.8, 80^\circ\}$ and $\{10^\circ, 0.92, 80^\circ\}$ with $-2$ min at MESSENGER and $-2$ min at STEREO-B. Negative values mean that ELEvoHI predicts the arrival earlier than observed. The ensemble mean of the prediction at MESSENGER is $\Delta t=-0.6 \pm 2.7$ h, the ensemble mean at STEREO-B is $\Delta t=-0.9 \pm 4.2$ h. The mean absolute error at MESSENGER is $\Delta t=2.2 \pm 1.6$ h and $\Delta t=3.5 \pm 2.6$ h at STEREO-B.
The ensemble median is $-0.21$ h at MESSENGER and $-0.03$ h at STEREO-B. The mean predicted arrival speed is $484 \pm 23$ km s$^{-1}$ at MESSENGER and $438 \pm 11$ km s$^{-1}$ at STEREO-B, while the in situ data show a speed variation in the sheath region between 350 and 400 km s$^{-1}$.


\subsection{Sensitivity analysis}

\begin{figure}[h]
\centering
\includegraphics[width=0.5\textwidth]{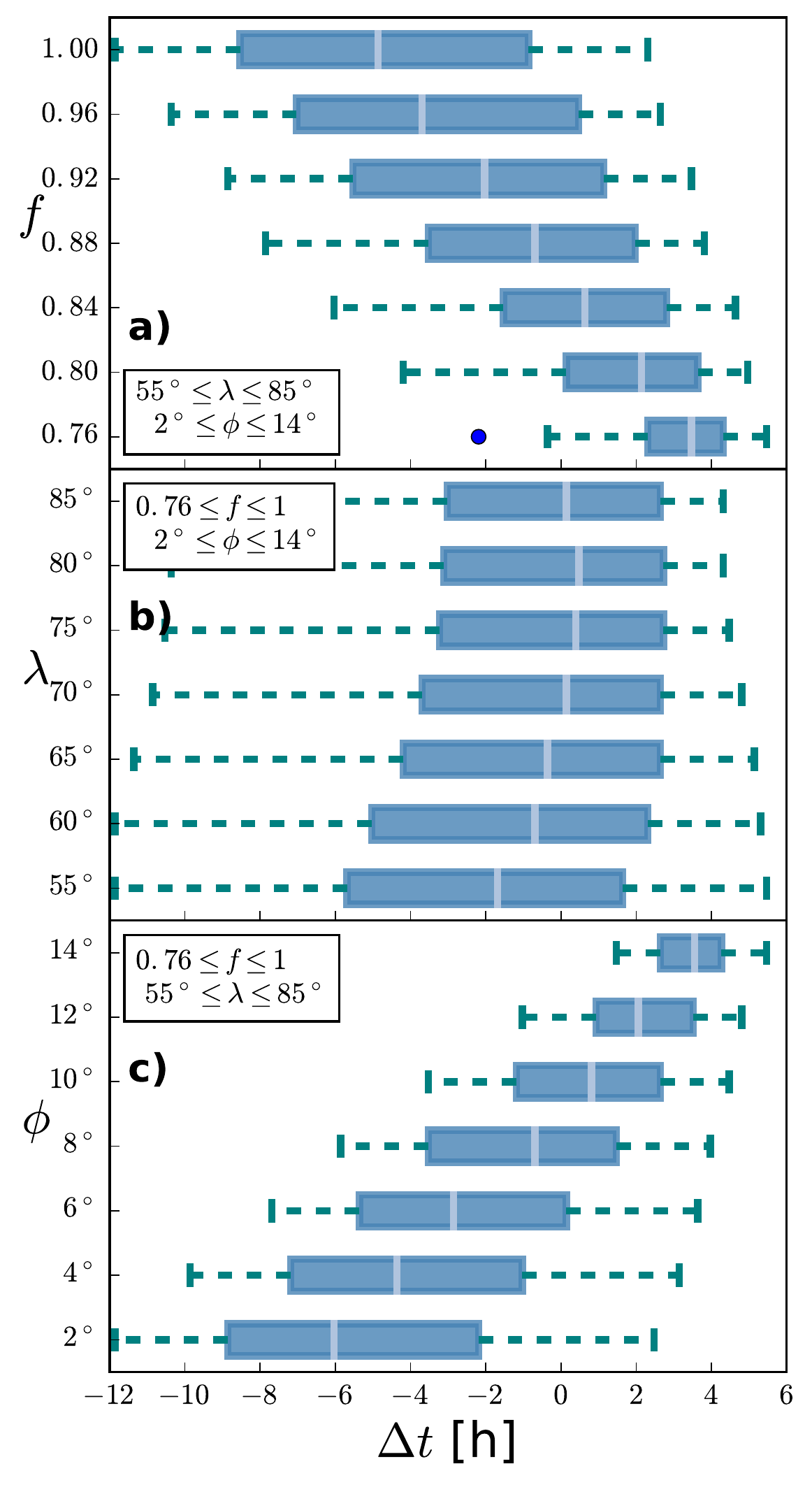}
\caption{Box and whiskers plots of $\Delta t$ for runs with one input parameter being fixed and the other two varying within their error ranges for each value of $f$, $\lambda$, and $\phi$ (a, b, c). The boxes encompass $50$\% of the data, with the vertical gray line representing the median. The whiskers extend out to $1.5$ times the interquartile range. The blue point in panel a) marks an outlier.}
\label{fig:dt_all}
\end{figure}

In order to test the robustness of ELEvoHI as a function of its shape-related input parameters, we examine the influence of each of the three input parameters by an analysis of the prediction variance. 
For each of the seven different values of each input parameter, the runs are arranged into groups. Box and whiskers plots for all groups for the different values of $f$ ($\lambda$, $\phi$) are displayed in Figure \ref{fig:dt_all}a (b, c). The $x$-axis shows the time difference between the predicted and observed arrival times, meaning that positive values correspond to an overestimated transit time.
The gray vertical lines mark the medians, the boxes encompass 50\% of the data, the whiskers extend out to $1.5$ times the interquartile range. The variance of the medians for the grouped results corresponding to fixed values of $f$ is $\sigma^2 = 7.9$ h ($\sigma=2.8$ h), while the median of all medians is $-0.7$ h. For the fixed values of $\lambda$ we find a variance of the medians of $\sigma^2 = 0.5$ h ($\sigma=0.7$ h), while the median of all medians is $0.1$ h and for the fixed values of $\phi$ the variance of the medians is $\sigma^2 = 10.5$ h ($\sigma=3.2$ h), and the median of all medians is $-0.7$ h.
In this case study, the highest influence on the prediction result has the direction of motion, meaning that a change of $\phi$ of $12^\circ$ leads to a difference in arrival time of $\approx 10$ h. In contrast, if an angular half width of $\lambda=55^\circ$ or $\lambda=85^\circ$ is used, only leads to a difference of 0.5 h.
However, it is important to note that this CME is propagating directly towards STEREO-B, which minimizes the influence of the CME shape on the prediction result. This may be different for events where not the CME apex is hitting the target of interest and it is likely to be of high importance when predicting flank encounters, where the CME width is a decisive factor if an impact is predicted or not. 

\section{Limiting the ensemble results}

An ensemble prediction is a great possibility to reveal the range of feasible prediction results and their occurrence frequencies. But are the most frequent predictions also the most accurate predictions? Is the mean or median value of the ensemble prediction a proper candidate to be used as resulting prediction? Is it possible to pin down the ensemble results to a most likely one? In this section, we explore the ensemble results to find a way to narrow down the forecasting range based on the CME mass or the occurrence frequencies of four resulting parameters, i.e.\ the drag parameter, the background solar wind speed and the initial distance and speed.

\subsection{Limitation using the CME mass}

In order to relate the CME mass derived from coronagraph observations to the ELEvoHI results, we now calculate the mass from ELEvoHI results using the definition of $\gamma$ from Equation \ref{eq:gamma} and rearrange

\begin{equation}
m_{\rm CME}=c_{\rm D} \frac{A \rho_{\rm sw}}{\gamma}.
\label{eq:mass_DBM}
\end{equation}

The cross section, $A$, is calculated assuming an ellipse perpendicular to the ecliptic plane, with the same semi-major axis, $a$, as resulting from ElCon. The semi-minor axis was calculated based on the angular width perpendicular to the ecliptic plane measured from the GCS fits, being 17$^\circ$ for the flux rope GCS fit and 21$^\circ$ for the GCS fit to the CME density pile-up. The angular width derived from GCS fitting is more reliable than using the apparent angular width from coronagraph observations, where the CME---due to projection effects---seems to be wider than it actually is \citep[][]{vrs07,wuche11}. The solar wind mass density, $\rho_{\rm sw}$, was calculated using the density model by \citet{leb98}, being simply a function of solar radial distance.

\begin{figure}[h]
\centering
\includegraphics[width=0.7\textwidth]{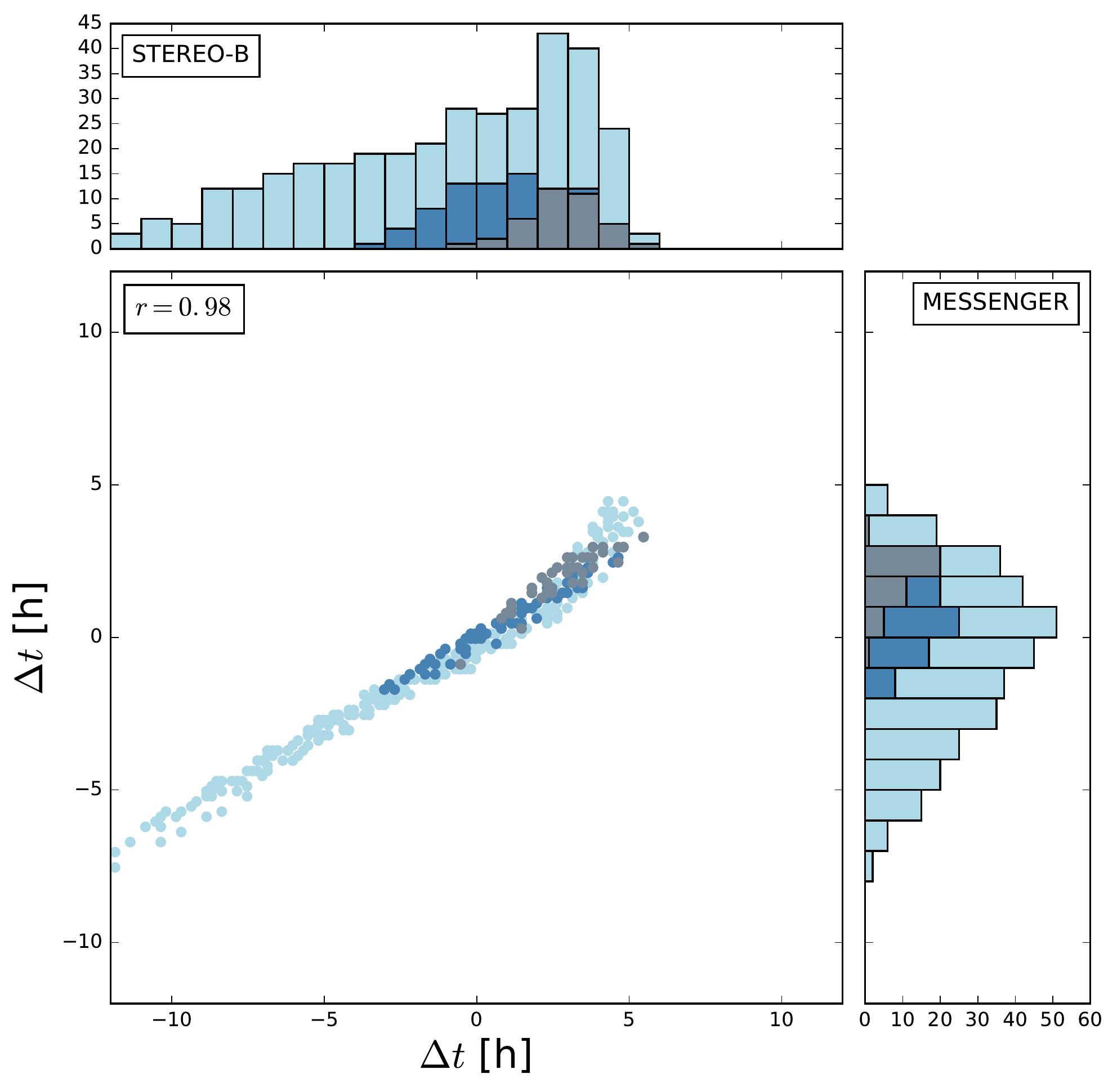}
\caption{Histogram of differences of observed and predicted arrival times, $\Delta t$, at STEREO-B (top) and MESSENGER (right). Positive values mean that the transit time is overestimated by ELEvoHI. The light blue bars show the distribution of all 339 runs, while the blue (gray) bars mark the runs with the calculated mass based on the wide GCS fit to the CME shock (narrow GCS fit to the flux rope) lying in a range of $\pm 20$\% of the mass calculated from coronagraph images. The correlation between $\Delta t$ at MESSENGER and STEREO-B is shown in the middle plot and yields a correlation coefficient of $r=0.98$.}
\label{fig:hist}
\end{figure}

The CME mass was calculated for each run at $r_{\rm init}$, located at $32 \pm 2.8$ R$_\odot$ on average. The CME mass derived from coronagraph observations at 30 R$_\odot$ ($m_{30}\approx 6.5~10^{15}$~g) is now used to verify parts of the ELEvoHI ensemble run. Figure \ref{fig:hist} shows a histogram of the differences between observed and predicted arrival times at STEREO-B (top panel) and MESSENGER (right panel) for the whole ensemble (light blue bars). The dark blue bars show the number of runs (81 events, i.e.\ 24\%), for which the masses derived from ELEvoHI based on the GCS fit to the CME density pile-up lie within $\pm 20$ \% of the mass calculated from coronagraph images, the gray bars represent the same based on the GCS flux rope fit (38 events, i.e.\ 11\%).
The mean arrival time difference of the sample based on the CME density pile-up is $1.1 \pm 1.9$ h at STEREO-B (compared to $\Delta t=-0.9 \pm 4.2$ h for the whole ensemble) and $0.7 \pm 1.2$ h at MESSENGER (compared to $\Delta t=-0.6 \pm 2.7$ h). The mean arrival time difference of the sample based on the CME flux rope is $2.8 \pm 1.3$ h at STEREO-B and $1.9 \pm 0.9$ h at MESSENGER. 
While the ELEvoHI mass based on the smaller angular width derived from the GCS flux rope fit leads to a larger difference between predicted and actual in situ arrival time, the ELEvoHI mass based on the GCS fit to the dense region surrounding the flux rope leads to a later prediction with almost the same difference to the in situ arrival. In the latter case it is possible to reduce the error range, i.e.\ the standard deviation, by about 50\% at STEREO-B and MESSENGER, respectively.
The scatter plot in Figure \ref{fig:hist} shows the correlation between $\Delta t$ at STEREO-B and MESSENGER for the whole ensemble (light blue) and for the mass constrained predictions (dark blue and gray). We find a correlation coefficient of $r=0.98$, meaning that an ELEvoHI run leading to a good prediction at $\approx 0.4$ AU also leads to a good prediction at 1 AU. However, this is not proven for events that hit the spacecraft with its flank or for not completely aligned spacecraft. Additionally, the correlation could be different if the CME has a higher speed and a higher drag parameter than the event under study or if the CME frontal shape is not in agreement with the elliptic assumption of ELEvoHI. However, there already are studies testing the ability of spacecraft located closer to the Sun along the Sun-Earth line to improve predictions, especially the prediction of the $B_z$ component of the magnetic flux rope within the CME \citep[e.g.][]{kub16}. 



\subsection{Limitation using $\gamma$, $w$, $r_{\mathrm{init}}$ and $v_{\mathrm{init}}$}

\begin{figure}[h]
 \centering
 \includegraphics[width=0.9\textwidth]{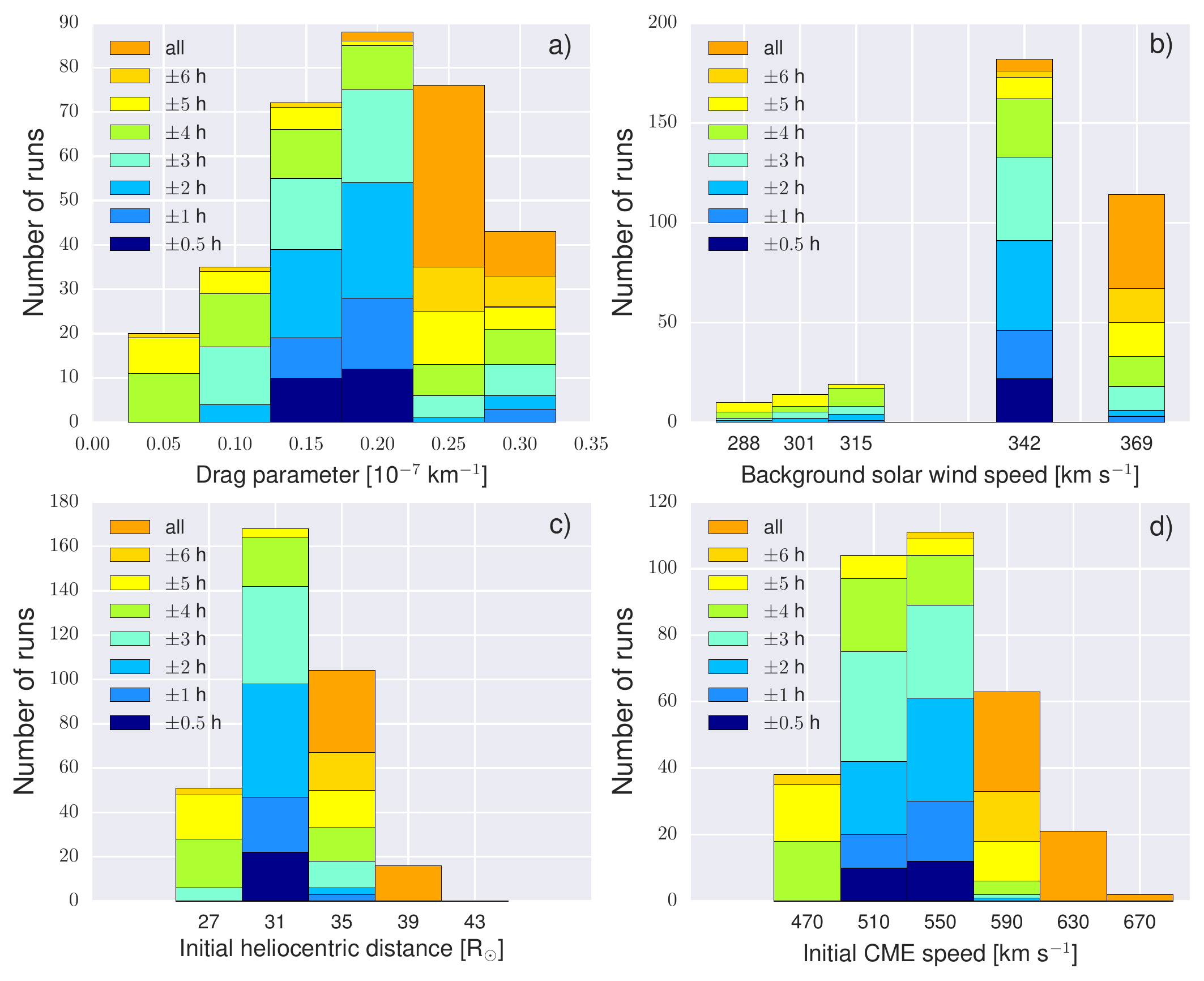}
 \caption{Frequency distribution of resulting drag parameter (a), $\gamma$, and background solar wind speed (b), $w$, initial heliocentric distance (c), and initial speed (d) resulting from all ELEvoHI runs and color-coded based on the predicted and observed arrival time differences.}
 \label{fig:gammaw}
 \end{figure}

ELEvoHI results cover the drag parameter, $\gamma$, and the background solar wind speed, $w$, both obtained by drag-based fitting to the ElCon time-distance profile. As described in Section \ref{sec:ELEvoHI}, the in situ solar wind speed from 1 AU from the same time-range as the HI observations are used to reveal the best candidate of $w$ for the fit. Five different fits are performed for five different values of $w$ within the minimum and the maximum values of the in situ solar wind speed. The fit with the smallest residuals reveals the resulting $w$. More information on this procedure can be found in \citet{rol16}. In contrast to $w$, $\gamma$ is in fact a true fitting result of the drag-based fit. Furthermore, we can gain information on $r_{\mathrm{init}}$ and $v_{\mathrm{init}}$ of the CME by ELEvoHI. In this model, $v_{\mathrm{init}}$ is derived from HI data after the conversion from elongation to distance. It depends on the chosen geometry of the CME front shape (within the ecliptic) and on the starting point of the DBM fit within the model. In this study, we fixed the starting point at the second measurement point in HI, i.e. the initial time ($t_{\mathrm{init}}$) is fixed---$r_{\mathrm{init}}$ and $v_{\mathrm{init}}$ are dependent on the shape and on $t_{\mathrm{init}}$.
This leads to a mean $r_{\mathrm{init}}$ of $32 \pm $3 R$_\odot$ ($r_{\mathrm{init, min}}=26.5$ and $r_{\mathrm{init, max}}=40.3$ R$_\odot$) for the whole ensemble.
The average $v_{\mathrm{init}}$ is $541 \pm 4$2 km s$^{-1}$ ($v_{\mathrm{init, min}}=460$ and $v_{\mathrm{init, max}}=661$ km s$^{-1}$).
Figure \ref{fig:gammaw}a shows the distribution of different values of $\gamma$, grouped in bins of a size of $0.05 \times 10^{-7}$ km$^{-1}$ and color-coded based on $\Delta t$. Surprisingly, all of the exact predictions (within $\pm 0.5$ h) and almost all predictions within $\pm 2$ h have a $\gamma$ of 0.15 or $0.2 \times 10^{-7}$ km$^{-1}$. Additionally, these values---along with $\gamma=0.25 \times 10^{-7}$ km$^{-1}$---are the most frequently resulting drag parameters in the whole ensemble. The same approach, but for $w$, $r_{\mathrm{init}}$ and $v_{\mathrm{init}}$ are presented in Figures \ref{fig:gammaw}b)-d). Here, we find the same picture: the best predictions result from the most frequent values. 

\begin{figure}[h]
 \centering
 \includegraphics[width=\textwidth]{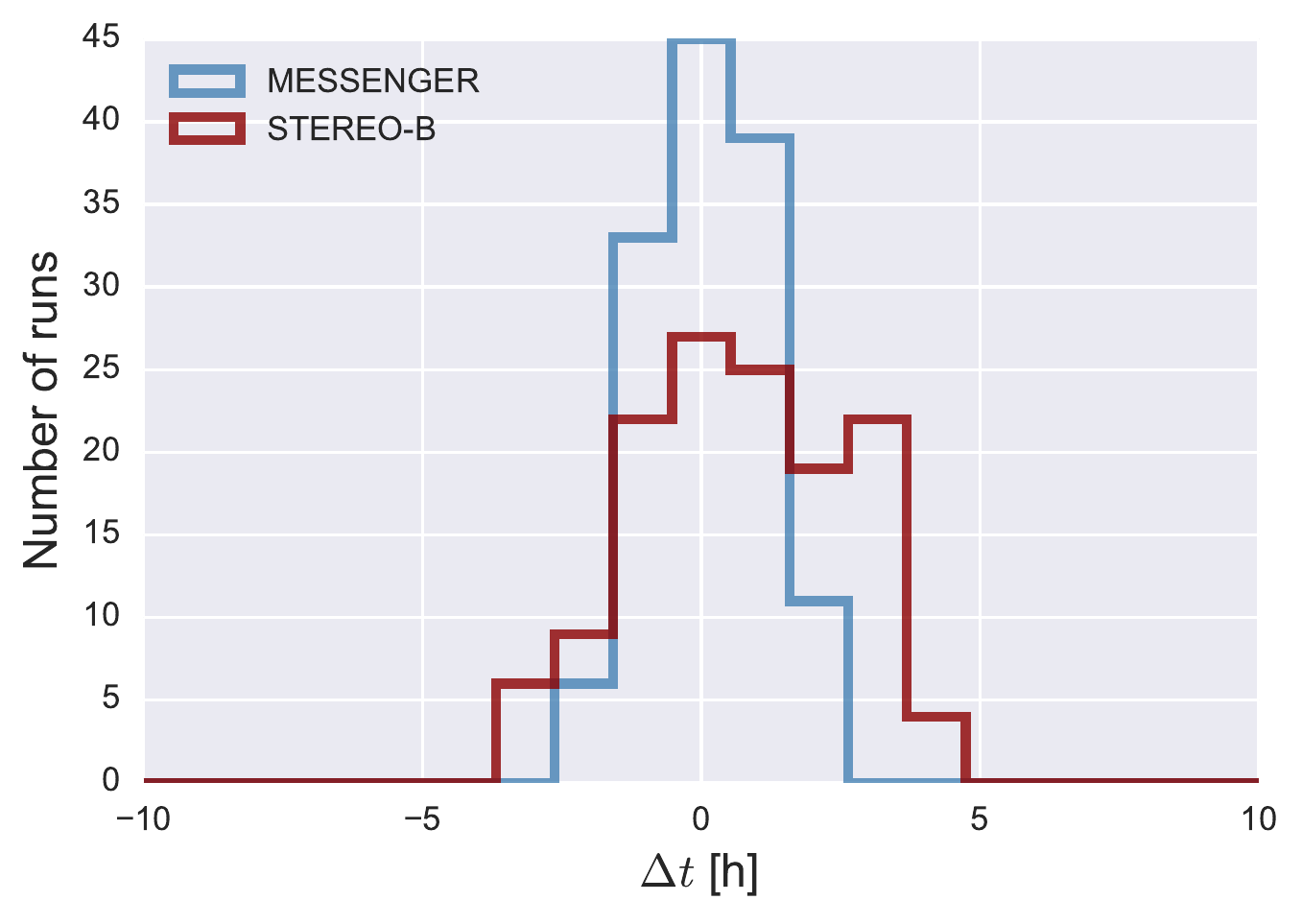}
 \caption{Frequency distribution of the predictions with $w=342$ km~s$^{-1}$ and $0.15 \times 10^{-7}$ km$^{-1} \leq \gamma \leq 0.25 \times  10^{-7}$ km$^{-1}$, $29$ R$_\odot \leq r_{\mathrm{init}} \leq 37$ R$_\odot$, and $490$ km s$^{-1} \leq v_{\mathrm{init}} \leq 570$ km s$^{-1}$, i.e.\ the most frequent values as seen in Figure \ref{fig:gammaw}.}
 \label{fig:wgammadt}
 \end{figure}

Since ELEvoHI is planned to be used as real-time prediction tool as soon as STEREO-A provides near real-time observations from the Sun-Earth line again, we try to find an approach to limit the ELEvoHI ensemble predictions, which can be used in real-time, i.e.\ it should be easy and fast.
When ensemble modeling is performed, taking into account the frequency distribution of, e.g.\ $\gamma$ and $w$ seems to be an easy and beneficial way to limit the ensemble results. As a proof of concept, we extract all runs where $0.15 \times 10^{-7}$ km$^{-1} \leq \gamma \leq 0.25 \times 10^{-7}$ km$^{-1}$ AND $w=342$ km s$^{-1}$, the runs with the most frequent values of $\gamma$ and $w$. This is the case for 140 runs. Additional limitation using the frequency distributions of $r_{\mathrm{init}}$ and $v_{\mathrm{init}}$ by taking into account only those runs with $29$ R$_\odot \leq r_{\mathrm{init}} \leq 37 $ R$_\odot$ and $490 $ km s$^{-1} \leq v_{\mathrm{init}} \leq 570 $ km s$^{-1}$ leads to a further reduction to 134 runs. Figure \ref{fig:wgammadt} shows the distribution of this sample and reveals that indeed, the predictions can be improved. In detail, the mean $\Delta t$ at MESSENGER is $0.1 \pm 1$ h (mean absolute error is $0.9 \pm 0.6$ h), the mean $\Delta t$ at STEREO-B is $0.7 \pm 1.8$ h (mean absolute error is $1.6 \pm 1.1$ h).

\begin{figure}[h]
 \centering
 \includegraphics[width=\textwidth]{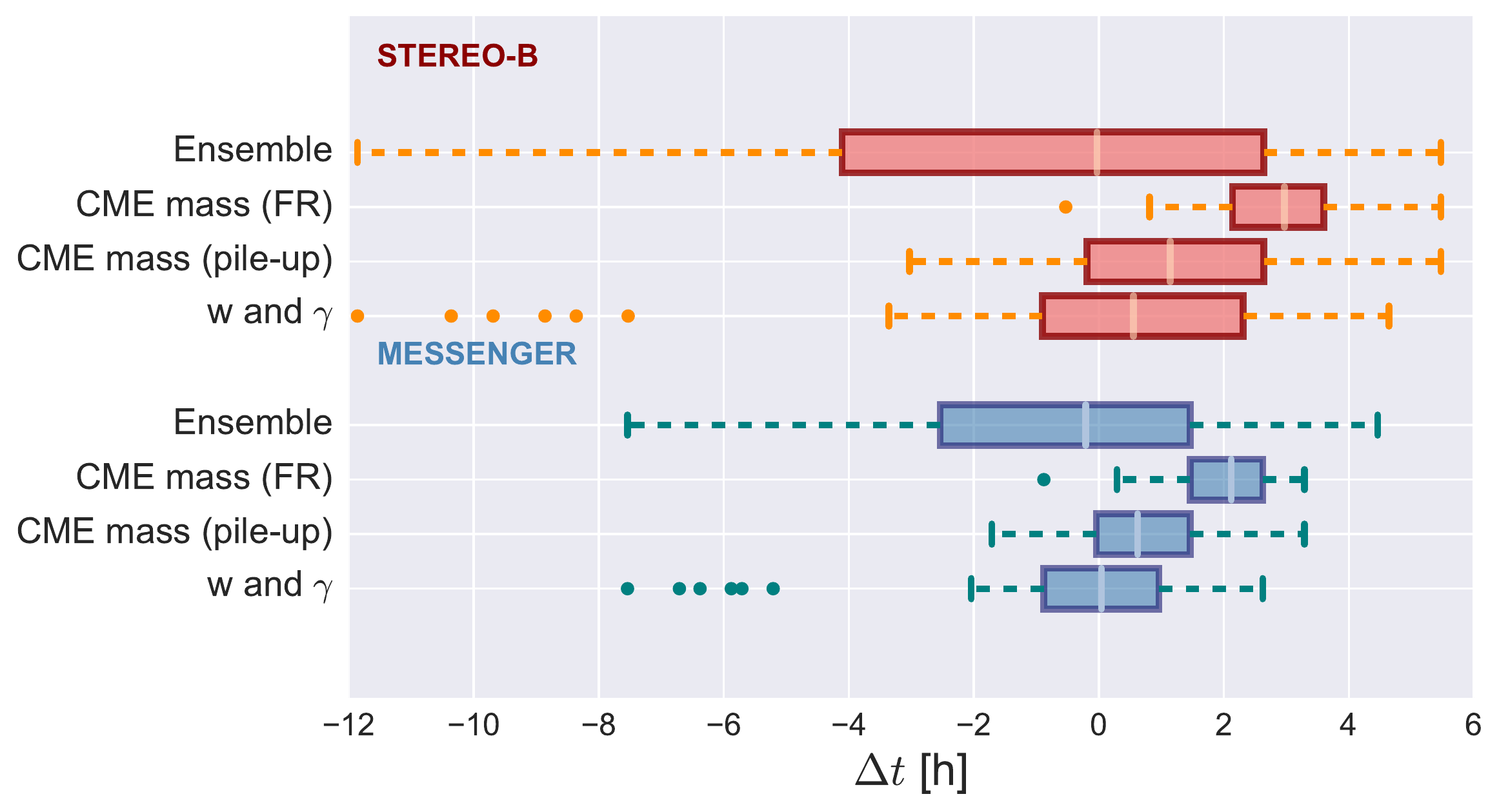}
 \caption{Comparison of the tested constraints to the ensemble results. Predictions for STEREO-B are red, predictions for MESSENGER are blue. In each case, the upper box represents the whole ensemble, the second corresponds to the mass-constrained sample based on the GCS flux rope fit (smaller angular width), the third corresponds to the mass-constrained sample based on the GCS fit to the pile-up region (larger angular width). The lower boxes show the ensemble resulting from the constraint of the most frequent values of $\gamma$ and $w$. The outliers (orange dots) would be excluded by further limiting the ensemble using the initial distance and speed distributions.}
 \label{fig:cons}
 \end{figure}

Figure \ref{fig:cons} presents a comparison of the two possibilities to limit the ELEvoHI ensemble prediction. The red boxes correspond to predictions for STEREO-B, the blue boxes correspond to predictions for MESSENGER. In each case, the upper box represents the whole ensemble, while the two middle boxes stand for the mass constrained sample and the lower box corresponds to the constraint using the most frequent values of $\gamma$ and $w$. The latter method is fast and simple, because no mass derivation is needed and the two parameters and their distribution directly result from the ensemble prediction. Furthermore, it is also more accurate than the mass constraint and can easily be used in real-time. The outliers (orange dots) can be excluded by further limiting the ensemble using the initial distance and speed distributions.

\subsection{Comparison to Enlil cone model}

To compare ELEvoHI with numerical model results on the interplanetary evolution of the shape of this CME we use the WSA--ENLIL+Cone model. The global 3D MHD ENLIL model provides a time-dependent description of the background solar wind plasma and magnetic field using the WSA coronal model \citep{argpiz00,arg04} as input at the inner boundary of 21.5 R$_{\odot}$ \citep{ods96,ods99a,ods99b,ods03,ods04}. A homogeneous, over-pressured hydrodynamic plasma cloud is launched through the inner boundary of the heliospheric computational domain and into the background solar wind. WSA coronal maps provide the magnetic field and solar wind speed at the boundary between the coronal and heliospheric models at 21.5\,R$_{\odot}$. ENLIL version 2.8 was used in this work, with a time-dependent inner boundary constructed from a series of daily input WSA synoptic maps, each computed from a new Global Oscillation Network Group \citep[GONG:][]{har96} daily synoptic ``QuickReduce'' magnetogram every 24 hours at the ENLIL inner boundary. For this study the WSA--ENLIL+Cone simulations have a 4$^{\circ}$ spatial resolution (low) and spherical grid size of 384$\times$30$\times$90 ($r,\theta,\phi$) with a three hour 3D output cadence and five minute output cadence at locations of interest. The simulation range was 0.1 to 2.1 AU in radius, $r$, $-60^{\circ}$ to $+60^{\circ}$ in latitude, $\theta$, and 0$^{\circ}$ to 360$^{\circ}$ in longitude, $\phi$. 

\begin{figure}[h]
\centering
\includegraphics[width=\textwidth]{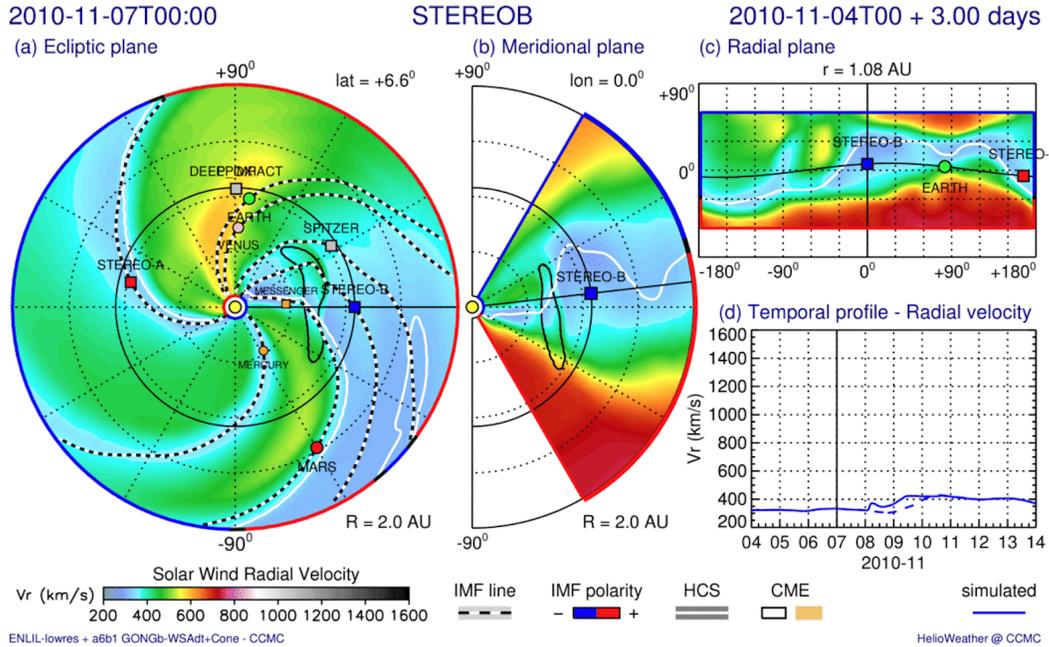}
\caption{Velocity contour plot of the CME simulation in the (a) ecliptic plane, (b) meridional plane of STEREO-B, and (c) 1 AU sphere in cylindrical projection on 7 November 2010 at 00:00 UT. Panel (d) shows the simulated (blue) radial velocity profile at STEREO-B.}
\label{fig:enlil}
\end{figure}

Figure \ref{fig:enlil} shows a velocity contour plot of the simulated CME in the ecliptic plane (a), the meridional plane of STEREO-B (b), a 1 AU sphere in cylindrical projection (c) and the simulated in situ solar wind speed at STEREO-B (d). The figure shows that nearly the center part of the CME impacts STEREO-B, followed by a high speed stream as seen in Figure \ref{fig:insitu}.
From the ENLIL run it can be seen how the simulated CME deformes during propagation. At its onset, the shape is almost elliptically, while during evolution a concave shape develops at the portion heading towards STEREO-B. One reason for this deformation may be the different drag regime on the other side of the heliospheric current sheet (white line) because of the higher solar wind speed (500 compared to 300 km s$^{-1}$). Interestingly, this dip in the CME front leads to an arrival time difference of 15 h compared to the arrival of a uniform shape. Here, the question arises if it is even possible to reduce today's real time prediction error to less than half a day.

\section{Summary and Discussion}

ELEvoHI is designed to predict CME arrivals in real-time provided that HI data are available in (near) real-time. To reach this goal until STEREO-A is close enough at the Sun-Earth line providing data, which can be used to predict Earth-directed CMEs, we assess and further develop this tool using science data. To predict CME arrivals with a high degree of accuracy it is of high importance to have suitable input parameters available. In this study we examined in which way GCS fitting to coronagraph observations is able to provide information on the shape of the CME front within the ecliptic plane and which influence the three shape-related input parameters (propagation direction, angular width, curvature of the front) have on the prediction result. 
Two different GCS fits were performed, namely one to the CME flux rope (as commonly done) and one capturing the dense area surrounding the flux rope. This latter GCS fit is assumed to be more consistent with other assumptions of ELEvoHI, especially with HI elongation measurements, which are taken at the shock front of the CME and not at the CME cavity. In order to identify the shape of the CME within the ecliptic, the GCS shape was subtended with the ecliptic plane resulting in an ellipse-shaped CME front. From this ellipse the needed input parameters and their error ranges were measured. 
Within this range of input parameters, an ensemble of 339 ELEvoHI runs was performed, predicting the arrival times and speeds at MESSENGER and STEREO-B. The ensemble mean for predictions at MESSENGER was $\Delta t=-0.6 \pm 2.7$ h and for STEREO-B $\Delta t=-0.9 \pm 4.2$ h. This is an impressing result, but one should keep in mind that case studies always lead to better predictions than studies dealing with larger event samples. Furthermore, HI science data are not available in real-time but were used in this study. If using HI beacon data leads to similar results needs to be further investigated in future studies. 

\subsection{L1 point as potential location for an HI observer}

The results of this study indicate that halo CMEs can be predicted by an HI observer at least as good as from aside. Of course, a study with a large event sample, where side and halo predictions are compared to each other is necessary to further investigate if a front view is maybe even better as a side view for the majority of events. Additionally, we used information about the shape from three viewpoints for the GCS fit, which are not be available from only one L1 observatory.
\citet{def16} already pointed out that the observation of Earth-directed CMEs may also be possible from the L1 point or in low Earth orbit (LEO), the latter location was already proposed and simulated by \citet{defhow15}. An operational space weather mission at L1 or LEO instead of L4 or L5 would reduce the costs of such a mission by a noteworthy amount.
\citet{moe17} found a false negative rate of 0.9 for self-predictions (i.e.\ remote and in situ observer is the same) using HI, which implies that HI at L1 might not be an appropriate space weather monitor. However, it is likely that the missing observations and predictions are due to the restricted field of view of HI, i.e.\ the cameras observe only one side of the Sun and miss events from the other side. Covering the eastern as well as the western Sun-Earth space with two separate HI observers pointing at opposite directions, may solve this issue. Additionally, \citet{moe17} used the same half width and curvature (namely circular) for all of the 1337 CMEs. The angular width of a CME is the key parameter deciding if a CME hits or misses the target of interest.

\subsection{Usage of ELEvoHI ensemble predictions}

In this study, we found a possibility to constrain the ELEvoHI ensemble prediction in a way that is easy and fast to conduct and leads to promising results for real-time predictions. Two different procedures were tested. The first approach is an exclusion of runs for which the mass resulting from ELEvoHI was not in agreement with the mass calculated from coronagraph observations. This approach is an additional verification of $\gamma$ resulting from ELEvoHI to avoid unphysical results. 
We calculated two different values for the cross-section area, on which the drag force is acting on, based on the angular width resulting from the GCS fitting to the CME flux rope and to the dense region preceding the flux rope, respectively. The cross-section area is needed to derive the mass from the ELEvoHI output. We found that mass calculated from the angular width derived from the GCS fit to the CME pile-up surrounding the flux rope reveals a better constraint of the ensemble results than the mass calculated from the more narrow GCS fit to the CME flux rope. Limiting the ensemble runs to those having the same mass ($\pm 20$\%) calculated from ELEvoHI as from coronagraph observations results in an error range of 50\% less ($\pm 2$ h) than for the whole ensemble. However, this might be a too slow approach for real time predictions but might be feasible if the procedure of the mass derivation from coronagraph observations is automatized.
For the second approach, the frequency distributions of $\gamma$, $w$, $r_{\mathrm init}$ and $v_{\mathrm init}$ resulting from the ensemble run (339 runs) showed that the most accurate predictions are connected to their most frequent values, resulting from drag-based fitting implemented within ELEvoHI. Taking into account only those runs where $\gamma$ as well as $w$, $r_{\mathrm init}$ and $v_{\mathrm init}$ belong to the most prevalent values, we were able to further constrain the ensemble prediction at MESSENGER to a mean error of $\Delta t=0.1 \pm 1$ h and at STEREO-B to $\Delta t=0.7 \pm 1.8$ h and to a mean absolute error of $\Delta t=0.9 \pm 0.6$ h at MESSENGER and to $\Delta t=1.6 \pm 1.1$ h at STEREO-B.

Ensemble forecasting seems to be a good possibility to use ELEvoHI for real-time prediction. A test by applying ELEvoHI to a large sample is going to reveal if this method is indeed an improvement or not. Furthermore, it may be worth testing if a GCS fit in advance of the ELEvoHI run can be avoided when doing an ensemble prediction. Varying the input parameters within their common values, e.g.\ $35^\circ \leq \lambda \leq 85^\circ$ and $0.4 \leq f \leq 1$, and extracting from the ensemble results the runs with the most frequent values of $\gamma$, $w$, $r_{\mathrm init}$ and $v_{\mathrm init}$ could speed up the prediction and make the usage of additional GCS fitting redundant.





\acknowledgments
Support by the Austrian Science Fund (FWF): P26174-N27 is acknowledged by TA and CM. The presented work has received funding from the European Union Seventh Framework Programme (FP7/2007-2013) under grant agreement No.~606692 [HELCATS]. This work was completed while PH held an NRC postdoctoral fellowship at the U.S.\ Naval Research Laboratory. MT acknowledges the support by the FFG/ASAP Programme under grant no.\ 859729 (SWAMI). MLM acknowledges the support of NASA LWS grant NNX15AB80G. This work is based in part on observations made with the Spitzer Space Telescope, which is operated by the Jet Propulsion Laboratory, California Institute of Technology under a contract with NASA. This research has made use of SunPy, an open-source and free community-developed solar data analysis package written in Python \citep[][]{mum15} and was carried out with the free Jupyter Anaconda environment. Simulation results have been provided by the Community Coordinated Modeling Center at the Goddard Space Flight Center through their public Runs on Request system (http://ccmc.gsfc.nasa.gov; run number Tanja\_Amerstorfer\_120516\_SH\_1). The WSA model was developed by N. Arge now at NASA/GSFC and the ENLIL Model was developed by D. Odstrcil now at GMU. We thank the \textit{STEREO SECCHI/IMPACT/PLASTIC} teams for their open data policy.

\bibliography{tam_bib}





\end{document}